\documentclass[twocolumn,english,superscriptaddress,longbibliography,reprint,prb]{revtex4-1}

\usepackage[T1]{fontenc}
\usepackage[latin9]{inputenc}
\usepackage{color}
\usepackage{bm}
\usepackage{amsmath}
\usepackage{amssymb}
\usepackage{graphicx}

\makeatletter
\usepackage{babel}

\makeatother

\usepackage{babel}
\begin{document}

\title{Neural-network quantum state tomography for many-body systems}

\author{Giacomo Torlai}

\affiliation{Department of Physics and Astronomy, University of Waterloo, Ontario
N2L 3G1, Canada}

\affiliation{Perimeter Institute of Theoretical Physics, Waterloo, Ontario N2L
2Y5, Canada}

\affiliation{Theoretische Physik, ETH Zurich, 8093 Zurich, Switzerland}

\author{Guglielmo Mazzola}

\affiliation{Theoretische Physik, ETH Zurich, 8093 Zurich, Switzerland}

\author{Juan Carrasquilla}

\affiliation{D-Wave Systems Inc., 3033 Beta Avenue, Burnaby BC V5G 4M9, Canada}

\author{Matthias Troyer}

\affiliation{Theoretische Physik, ETH Zurich, 8093 Zurich, Switzerland}

\affiliation{Quantum Architectures and Computation Group, Station Q, Microsoft
Research, Redmond, WA 98052, USA}

\author{Roger Melko}

\affiliation{Department of Physics and Astronomy, University of Waterloo, Ontario
N2L 3G1, Canada}

\affiliation{Perimeter Institute of Theoretical Physics, Waterloo, Ontario N2L
2Y5, Canada}

\author{Giuseppe Carleo}

\affiliation{Theoretische Physik, ETH Zurich, 8093 Zurich, Switzerland}
\begin{abstract}
The experimental realization of increasingly complex synthetic quantum
systems calls for the development of general theoretical methods,
to validate and fully exploit quantum resources. Quantum-state tomography
(QST) aims at reconstructing the full quantum state from simple measurements,
and therefore provides a key tool to obtain reliable analytics. Brute-force
approaches to QST, however, demand resources growing exponentially
with the number of constituents, making it unfeasible except for small
systems. Here we show that machine learning techniques can be efficiently
used for QST of highly-entangled states, in both one and two dimensions.
Remarkably, the resulting approach allows one to reconstruct traditionally
challenging many-body quantities \textendash{} such as the entanglement
entropy \textendash{} from simple, experimentally accessible measurements.
This approach can benefit existing and future generations of devices
ranging from quantum computers to ultra-cold atom quantum simulators. 
\end{abstract}
\maketitle
Machine-learning (ML) methods have been demonstrated to be particularly
powerful at compressing high-dimensional data into low-dimensional
representations. \cite{hinton_reducing_2006,lecun_deep_2015} Thanks
to its intrinsic flexibility, ML is being applied to unravel complex
patterns hidden in the most diverse data sources, showing robustness
against noise, and receptiveness to generalization. While in the past
ML has been mostly applied to data science, it has recently been used
to address questions in the physical sciences. Applications to quantum
many-body systems have been put forward last year, to classify phases
of matter \cite{wang_discovering_2016,carrasquilla_machine_2017,broecker_machine_2016,van_nieuwenburg_learning_2017},
and to improve the simulation of classical \cite{huang_accelerated_2017,liu_self-learning_2017}
and quantum \cite{carleo_solving_2016} systems.

QST is by itself a data-driven problem, in which we aim to obtain
a complete quantum-mechanical description of a system, on the basis
of a limited set of experimentally accessible measurements.\cite{vogel1989determination}
Key quantum features, such as multi-qubit entanglement, are however
challenging to probe directly in current experimental setups. \cite{islam_measuring_2015}
Finding an efficient method to reliably extract such information from
a generic quantum device is therefore important for the development
of more powerful quantum simulators. In order to efficiently perform
QST, it is necessary to find a compact, and sufficiently general representation
of the quantum state to be analyzed. Matrix-Product-States (MPS) is
certainly the state-of-the-art tool for the tomography of low-entangled
states \cite{cramer2009efficient,lanyon2016efficient}, but alternative
representations are urged when performing QST of highly-entangled
quantum states, resulting either from deep quantum circuits or high-dimensional
physical systems.

In this Letter we show how ML approaches can be used to find such
representations. In particular, we argue that suitably-trained artificial
neural networks (ANN) offer a natural, efficient, and general way
of performing QST driven by experimental data. Our approach is demonstrated
on controlled artificial datasets, comprising measurements from several
quantum states with a large number of degrees of freedom (qubits,
spins, etc...), that are thus traditionally hard for QST approaches.

The ANN architecture we use in this work is based on restricted Boltzmann
machine (RBM) models. RBMs feature a visible layer (describing the
physical qubits) and a hidden layer of stochastic binary neurons fully
connected with weighted edges to the visible layer. These models have
been successfully employed to effectively solve complex many-body
problems.\cite{carleo_solving_2016,torlai_learning_2016,torlai_neural_2016}
``Neural quantum state'' representations of the many-body wave-function
have been shown to be capable of sustaining high entanglement, and
to efficiently describe complex topological phases of matter.\cite{carrasquilla_machine_2017,deng_exact_2016,deng_quantum_2017,gao_efficient_2017,chen_equivalence_2017,huang_neural_2017}
Given these favorable properties, RBM-based quantum states are natural
candidates for QST of low and high-dimensional many-body systems.

Let us consider, given some reference basis $\bm{x}$ (e.g. $\bm{\sigma}^{z}$
for spin-$\frac{1}{2}$), a generic many body target wave-function
$\Psi(\bm{x})\equiv\langle\bm{x}|\Psi\rangle$ describing the physical
system of interest. We introduce then an RBM wave-function: 
\begin{equation}
\psi_{\bm{\lambda},\bm{\mu}}(\bm{x})=\sqrt{\frac{p_{\bm{\lambda}}(\bm{x})}{Z_{\bm{\lambda}}}}\:\text{e}^{i\phi_{\bm{\mu}}(\bm{x})/2}
\end{equation}
where $Z_{\bm{\lambda}}$ is the normalization constant, $\phi_{\bm{\mu}}=\log p_{\bm{\mu}}(\bm{x})$
and $p_{\bm{\lambda}/\bm{\mu}}(\bm{x})$ are RBM probability distributions
corresponding to two different sets $\bm{\lambda}/\bm{\mu}$ of network
parameters (see Suppl. Inf.). Our ML approach to QST is then carried
out as follows. First, the RBM is trained on a dataset consisting
of a series of independent density measurements $|\Psi(\bm{x}^{[b]})|^{2}$
realized in a collection of bases $\{\bm{x}^{[b]}\}$ of the $N$-body
quantum system. During this stage, the network parameters $(\bm{\lambda},\bm{\mu})$
are optimized to maximize the dataset likelihood, in a way that $|\psi_{\bm{\lambda},\bm{\mu}}(\bm{x}^{[b]})|^{2}\simeq|\Psi(\bm{x}^{[b]})|^{2}$
(see Suppl. Inf.). Once trained, $\psi_{\bm{\lambda},\bm{\mu}}(\bm{x})$
approximates both the wave-function's amplitudes and phases, thus
reconstructing the target state. The accuracy of the reconstruction
can be systematically improved by increasing the number of hidden
neurons $M$ in the RBM for fixed $N$, or equivalently the density
of hidden units $\alpha=M/N$.\cite{carleo_solving_2016} One key
feature of our QST approach, is that it only needs raw data, i.e.
many experimental snapshots coming from single measurements, rather
than estimates of expectation values of operators.\cite{vogel1989determination,haffner_scalable_2005,cramer2009efficient,lanyon2016efficient,lanyon_efficient_2016,Toth2010,lu_experimental_2007}
This setup implies that we circumvent the need to achieve low levels
of intrinsic Gaussian noise in the evaluations of mean values of operators.

\begin{figure}[t]
\noindent \centering{}\includegraphics[width=1\columnwidth]{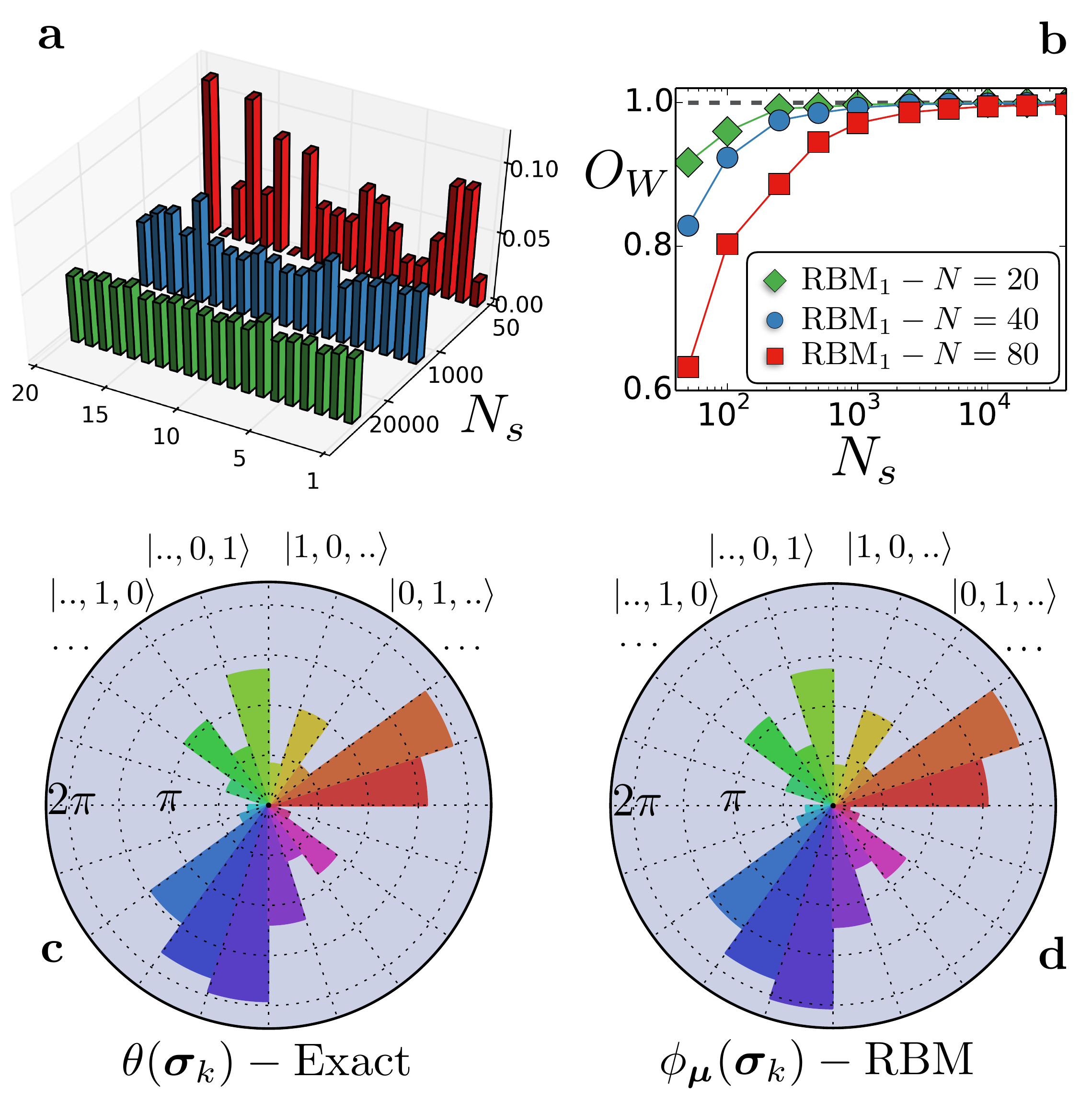}
\caption{\textbf{Tomography of the $W$ state.} \textbf{a)} Overlap between
the $W$ state and the wave-function generated by the trained RBM
with $\alpha=1$ as a function of the number of samples in the training
dataset. \textbf{b)} Histogram of the occurrence of each of the superposed
states in the $W$ state for $N=20$ qubits. We plot three histograms
obtained by sampling a RBM trained on a dataset containing $50$ (red),
$1000$ (blue) and $20000$ (green) independent samples. \textbf{c-d)}
Phases $\theta(\bm{\sigma}_{k}^{z})$ for each of the $N=20$ states
(different colors) in the phase augmented W state. We show the comparison
between the exact phases (\textbf{c}) and the phases learned by a
RBM (\textbf{d}), trained using 6400 samples per basis (magnitudes
of the phases are plotted along the radial direction). RBM tomography
allows here to systematically converge to the target $W$ state for
both cases with real and complex wave-function coefficients, upon
increasing the number of experimental samples.}
\label{Figure1} 
\end{figure}

To demonstrate the power of this approach, we start by considering
QST of the $W$ state, a paradigmatic $N$-qubit multipartite entangled
wave-function defined as 
\begin{equation}
|\Psi_{W}\rangle=\frac{1}{\sqrt{N}}\big(|100\dots\rangle+
...+|\dots001\rangle\big).\label{eq:WState}
\end{equation}
To mimic experiments, we generate several datasets with an increasing
number of synthetic density measurements obtained by sampling from
the $W$ state in the $\bm{\sigma}^{z}$ basis. These measurements
are used to train an RBM model featuring only the set of parameters
$\bm{\lambda}$, since the target $|\Psi_{W}\rangle$ is real and
positive in this basis. After the training, we sample from $|\psi_{\bm{\lambda}}(\bm{\sigma}^{z})|^{2}$
and build the histogram of the frequency of the components appearing
in $|\Psi_{W}\rangle$. In Fig.~\ref{Figure1}(\textbf{a}) we show
three histograms obtained with a different number of samples in the
training dataset for $N=20$, and for a fixed density of hidden variables
$\alpha=1$. From the histograms, we see that upon increasing the
number of samples each of the $N$ components $\big(|100\dots\rangle,|010\dots\rangle\dots)$
contribute equally to the wave-function, as expected from the exact
$W$ state. To better quantify the quality of our reconstruction we
then compute the overlap $O_{W}=|\langle\Psi_{W}|\psi_{\bm{\lambda}}\rangle|$
of the wave-function generated by the RBM with the original $W$ state
(see Suppl. Mat.). In Fig.~\ref{Figure1}(\textbf{b}) $O_{W}$ is
shown as a function of the number of samples in the training datasets
for three different values of $N$. For a system size substantially
larger than what is currently available in experiments,\cite{wang_experimental_2016}
an overlap $O_{W}\sim1$ can be achieved with a limited number of
samples. As a comparison, for $N=8$, full QST requires almost $10^{6}$
measurements,\cite{haffner_scalable_2005} whereas our approach achieves
comparable accuracy with only about $100$ measurements. We further
consider a phase-augmented W state, where a local phase shift $\text{exp}(i\theta(\bm{\sigma}_{k}^{z})/2)$
with random phase $\theta(\bm{\sigma}_{k}^{z})$ is applied to each
qubit. QST is now carried out using the full RBM wave-function and
training on $2(N-1)$ additional bases (see Suppl. Mat.). In the lower
section of Fig.~\ref{Figure1} we plot the comparison between the
exact phases (\textbf{c}) and the phases learned by the RBM (\textbf{d})
for $N=20$ qubits, showing very good agreement ($O_{W}=0.997$).

We now turn to the case of more complex systems and demonstrate QST
for genuine many-body problems. To mimic experimental outcomes, we
generate artificial datasets sampling different quantum states of
interacting spin models on a lattice. These are directly relevant
for quantum simulators based on ultra-cold ions and atoms. In particular
we consider the transverse-field Ising model (TFIM) with Hamiltonian
\begin{equation}
\mathcal{H}=\sum_{ij}J_{ij}\sigma_{i}^{z}\sigma_{j}^{z}-h\sum_{i}\sigma_{i}^{x}\label{eq:TFIM}
\end{equation}
and the XXZ spin-$\frac{1}{2}$ model, with Hamiltonian 
\begin{equation}
\mathcal{H}=\sum_{ij}\bigg[\Delta\left(\sigma_{i}^{x}\sigma_{j}^{x}+\sigma_{i}^{y}\sigma_{j}^{y}\right)+\sigma_{i}^{z}\sigma_{j}^{z}\bigg]\label{eq:XXZ}
\end{equation}
where the ${\bf \sigma}_{i}$ are Pauli spin operators.

\begin{figure*}[t]
\noindent \centering{}\includegraphics[width=2\columnwidth]{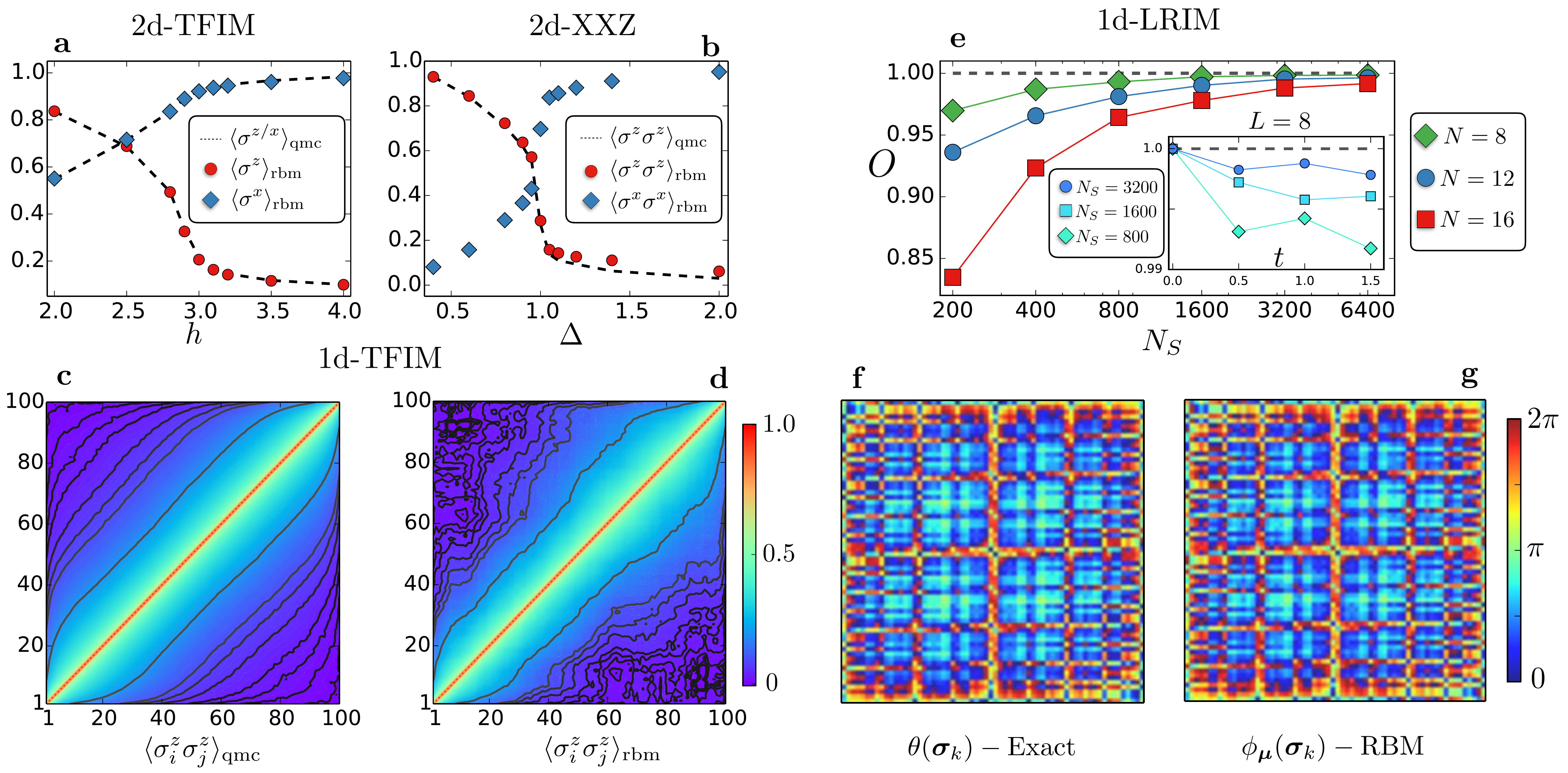}
\caption{\textbf{Tomography for many-body Hamiltonians.} In panels \textbf{(a-d)}
we show QST for ground states, comparing the reconstructed observables
to those obtained with quantum Monte Carlo simulations. In panels
\textbf{(e-g)} we show QST for unitary evolution of a 1d chain following
a quantum quench with long-range Ising Hamiltonian with $\gamma=3/4$.
\textbf{a)} Diagonal and off-diagonal magnetizations as a function
of the transverse field $h$ for the ferromagnetic 2d-TFIM on a square
lattice with linear size $L=12$ ($N=144$). \textbf{b)} Two-point
correlation function (diagonal and off-diagonal) between neighboring
spins along the diagonal of the square lattice (linear size $L=12$)
for the 2d-XXZ model. Each data point is obtained with a RBM from
a network trained with $\alpha=1/4$ on separate datasets. RBM-QST
allows here to accurately reconstruct, for each model, both diagonal
and off-diagonal observables of the target state. In the lower panels
we show the reconstruction of the diagonal spin correlation function
$\langle\sigma_{i}^{z}\sigma_{j}^{z}\rangle$ for the 1d-TFIM with
$N=100$ sites at the critical point $h=1$. \textbf{c)} Direct calculation
on spin configurations from a test-set much larger than the training
dataset, \textbf{d)} Reconstruction of the correlations by sampling
the trained RBM with $\alpha=1/2$. \textbf{e)} Overlap between the
system wave-function $\Psi(\bm{\sigma};t)$ and the RBM wave-function
$\psi_{\bm{\lambda},\bm{\mu}}(\bm{\sigma})$ for $t=0.5$, as a function
of the number of samples $N_{S}$ per basis. In the inset we show
the overlap as a function of time for different values of $N_{S}$.
In the lower panels we show the reconstruction of the $2^{N}$ phases
(re-arranged as a 2d array) for $N=12$ and $t=0.5$. \textbf{f)}
Exact phases $\theta(\bm{\sigma}_{k})$ for each component $\Psi(\bm{\sigma}_{k};t)$.
\textbf{g)} Phases $\phi_{\bm{\mu}}(\bm{\sigma}_{k})$ learned by
the RBM with $\alpha=1$.}
\label{many_body_H} 
\end{figure*}

We first discuss QST for ground state wave-functions of Hamiltonians
with nearest neighbors couplings, considering both a 1-dimensional
(1d) chain with $N$ sites and a 2-dimensional (2d) square lattice
with linear extent $L$ (for a total of $N=L^{2}$ spins). Synthetic
measurements in this case are obtained with standard quantum Monte
Carlo (QMC) methods (see Supp. Inf.), stochastically sampling the
exact ground-state of Hamiltonians in Eqs. (\ref{eq:TFIM},\ref{eq:XXZ})
for different values of the coupling parameters $h$ and $\Delta$,
covering the critical part of the phase diagram. The many-body ground-state
wave-function is real and positive, thus our reconstruction scheme
does not require measurements in any additional basis other than $\bm{\sigma}^{z}$.
Once the training is complete, we can test the representational power
of the neural networks by computing various observables using the
RBM and comparing with the values obtained through QMC simulations.\cite{torlai_learning_2016}.
In particular we consider few-body magnetic observables, such as magnetization
and spin correlations.

For the TFIM we look both at the longitudinal $\sigma^{z}$, and transverse
$\sigma^{x}$ magnetizations. As shown in Fig.~\ref{many_body_H}
(\textbf{a}) for $d=2$, the RBMs can reproduce the average values
with high accuracy, both for diagonal and off-diagonal observables.
For the XXZ model, we show in Fig.~\ref{many_body_H} (\textbf{b})
for $d=2$ the expectation values of the diagonal $\sigma_{a}^{z}\sigma_{b}^{z}$
and off-diagonal $\sigma_{a}^{x}\sigma_{b}^{x}$ spin correlations,
with $a$ and $b$ being neighbors along the lattice diagonal. Finally,
we consider the full spin-spin $\sigma_{i}^{z}\sigma_{j}^{z}$ correlation
function for the 1d-TFIM, which involves non-local correlations. We
show the reconstruction of the correlation function using the RBM
(Fig.~\ref{many_body_H} (\textbf{d})) closely matching the exact
result obtained by direct computation from the spin states on a much
larger independent set of QMC measurements (Fig.~\ref{many_body_H}
(\textbf{c})), with deviations compatible with statistical uncertainty
due to the finiteness of the training set.

In the context of many-body Hamiltonians, we now go beyond ground
states and realize QST for states originating from dynamics under
unitary evolution. In particular, we consider a 1d chain of Ising
spins initially prepared in the state $\Psi_{0}=|\rightarrow,\rightarrow,\dots,\rightarrow\rangle$
(fully aligned in the $\bm{\sigma}^{x}$ basis), subject to unitary
dynamics enforced by the Hamiltonian in Eq.~\ref{eq:TFIM} with long-range
interactions $J_{ij}\propto1/|i-j|^{\gamma}$ and magnetic field set
to zero ($h=0$). This kind of ``quench'' dynamics is realizable
in experiments with ultra-cold ions\cite{Richerme:2014aa}. For a
given time $t$, we perform QST on the state $|\Psi(t)\rangle=\text{exp}(-i\mathcal{H}t)|\Psi_{0}\rangle$
by training the RBM on spin density measurements performed in $2N+1$
different basis (see Supp. Inf.). In Fig.~\ref{many_body_H} (\textbf{e})
we show the overlap between the RBM wave-function $\psi_{\bm{\lambda},\bm{\mu}}(\bm{\sigma})$
and the time-evolved state $\Psi(\bm{\sigma};t)$ for different system
sizes $N$, as a function of the number $N_{S}$ of samples per basis
(inset shows overlap scaling with time). In the lower plot we show
for $N=12$ the exact (\textbf{f}) and the reconstructed phases (\textbf{g}).
The quality of the RBM-QST is once more remarkable, with a limited
number of measurements needed.

\begin{figure}
\noindent \centering{}\includegraphics[width=0.8\columnwidth]{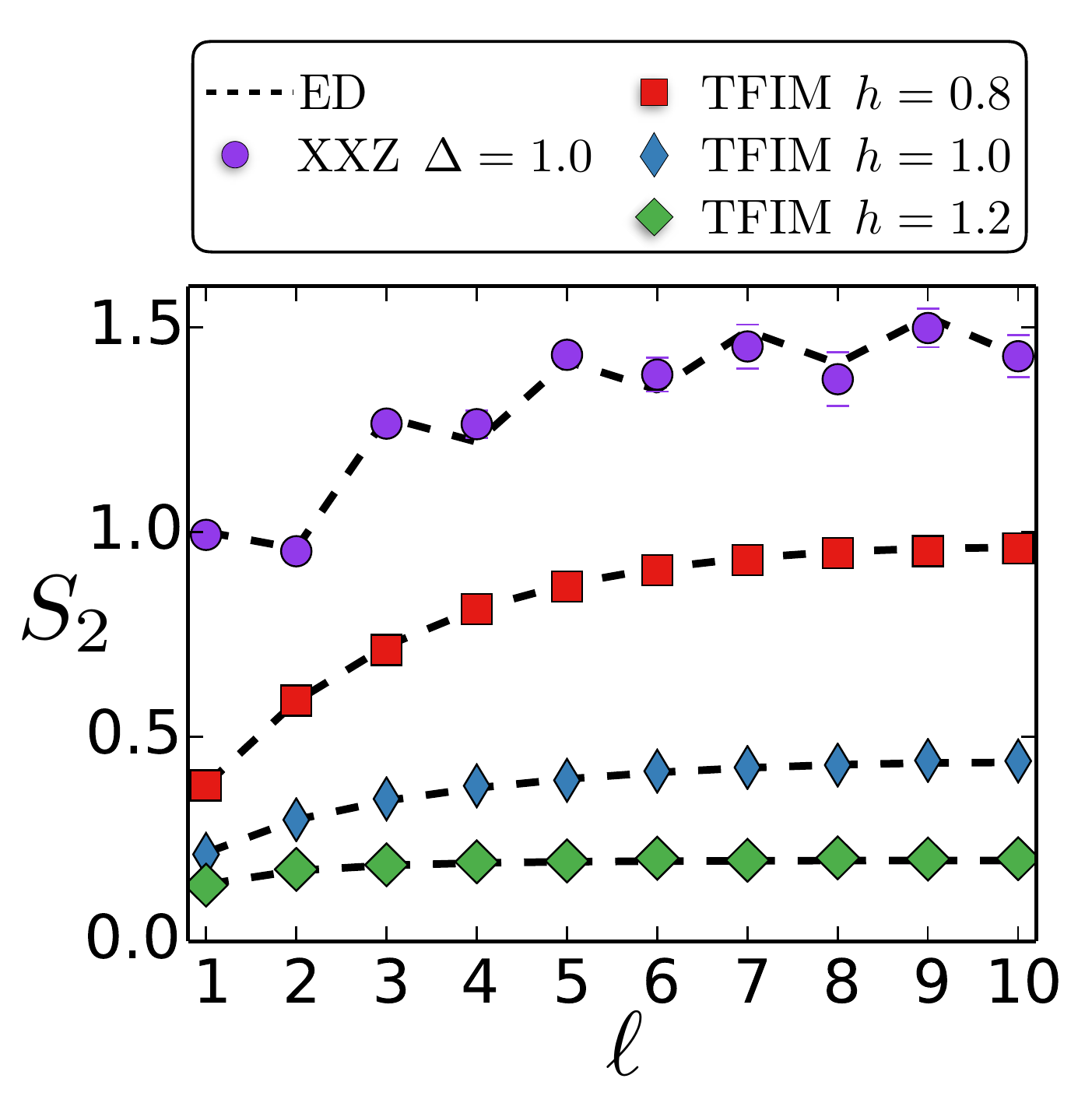}
\caption{\textbf{Reconstruction of the entanglement entropy.} Second Renyi
entropy as a function of the subsystem size $\ell$ for $N=20$ spins.
We compare results obtained using the the RBM wave-functions (markers)
with exact diagonalization (dashed lines) for the 1d-TFIM at different
values of the transverse magnetic field $h$ and the 1d-XXZ model
with critical anisotropy $\Delta=1$.}
\label{entropy} 
\end{figure}

To further assess the capabilities of our approach, we finally turn
to the entanglement entropy, a highly non-local quantity particularly
challenging for direct experimental observations.\cite{islam_measuring_2015}
It provides important information on the universal behavior of interacting
many-body systems and it is of central interest in condensed matter
physics and quantum information theory. Following the method proposed
here, we can obtain an estimate of this quantity given only simple
measurements of the density, which are more accessible with current
experimental advances. \cite{bakr_probing_2010} Given a bipartition
of the physical system, we consider in particular the second Renyi
entropy defined as $S_{2}(\rho_{A})=-\log(\mbox{Tr}(\rho_{A}^{2}))$,
with the subsystem $\rho_{A}$ of varying size. We estimate $S_{2}$
by employing an improved ratio trick sampling~\cite{hastings_measuring_2010}
using the wave-function generated by the RBM. In Fig.~\ref{entropy}
we show the entanglement entropy for the 1d-TFIM with three values
of the transverse field, and for the critical ($\Delta=1$) 1d-XXZ
model. In both instances we took a chain with $N=20$ spins and plot
the entanglement entropy as a function of the subsystem size $\ell\in[1,N/2]$.
The values obtained with the RBM (markers) are compared with results
from exact diagonalization (dashed lines), with an overall good agreement.

To conclude, we have demonstrated that ML tools can be efficiently
used to reconstruct complex many-body quantum states from a limited
number of experimental measurements. Our scheme is general enough
to be efficiently applied to a variety of quantum devices for which
current approaches demand exponentially large resources. These include
QST of highly-entangled quantum circuits, adiabatic quantum simulators,\cite{johnson_quantum_2011}
experiments with ultra-cold atoms and ions traps in higher dimensions.\cite{bloch_many-body_2008,blatt_quantum_2012,shulman_demonstration_2012}
Our approach can be used to directly validate quantum computers and
simulators, as well as to indirectly reconstruct quantities which
are experimentally challenging for a direct observation. For example,
we anticipate that the current generation of quantum microscopes could
substantially benefit from neural-quantum states QST. In particular,
we predict that the use of our approach for bosonic ultra-cold atoms
experiments would allow for the determination of the entanglement
entropy on systems substantially larger than those currently accessible
with quantum interference techniques.\textcolor{blue}{\cite{islam_measuring_2015} }

\subsection*{Acknowledgements }

We thank H. Carteret and B. Kulchytskyy for useful discussions. GT
thanks the Institute for Theoretical Physics, ETH Zurich, for hospitality
during various stages of this work. GT and RGM acknowledge support
from NSERC, the Canada Research Chair program, the Ontario Trillium
Foundation, and the Perimeter Institute for Theoretical Physics. Research
at Perimeter Institute is supported through Industry Canada and by
the Province of Ontario through the Ministry of Research \& Innovation.
GC, GM and MT acknowledge support from the European Research Council
through ERC Advanced Grant SIMCOFE, and the Swiss National Science
Foundation through NCCR QSIT. Simulations were performed on resources
provided by SHARCNET, and by the Swiss National Supercomputing Centre
CSCS.

\clearpage
\appendix
\onecolumngrid

\section{RBM Quantum State Tomography}

We provide in this section a description of the different steps required
to perform quantum state tomography (QST) with neural networks for
many-body quantum systems. We concentrate on the case of systems with
two local degrees of freedom (spin-$\frac{1}{2}$, qubits, etc) and
choose $\bm{\sigma}\equiv\bm{\sigma}^{z}$ as the reference basis
for the $N$-body wave-function $\Psi(\bm{\sigma})\equiv\langle\bm{\sigma}|\Psi\rangle$
we intend to reconstruct. This high-dimensional function can be well
approximated with an artificial neural network (NN). Given a set of
input variables (for example $\bm{\sigma}=\sigma_{1},\sigma_{2},\dots,\sigma_{N}$),
a NN is a highly non-linear function whose output is determined by
some internal parameters $\bm{\kappa}$. The architecture of the network
consists of a collection of elementary units, called neurons, connected
by weighted edges. The strength of these connections, specified by
the parameters $\bm{\kappa}$, encode conditional dependence among
neurons, in turn leading to complex correlations among the input variables.
Increasing the number of auxiliary neurons systematically improves
the expressive power of the NN function, which can then be used as
a general-purpose approximator for the target wave-function.\cite{carleo_solving_2016}
Goal of our scheme, is to find the best NN approximation for the many-body
wave-function, $\psi_{\bm{\kappa}}(\bm{\sigma})$, using only experimentally
accessible information.

The QST scheme proposed proceeds as follows. First, we assume that
a set of experimental measurements in a collection of bases $b=0,1,2\dots N_{B}$
is available. These measurements are distributed according to the
probabilities $P_{b}(\bm{\sigma}^{[b]})\propto|\Psi(\bm{\sigma}^{[b]})|^{2}$,
thus contain information about both the amplitudes and the phases
of the wave-function in the reference basis $\bm{\sigma}$. Goal of
the NN training, is to find the optimal set of parameters $\bm{\kappa}$
such that $\psi_{\bm{\kappa}}(\bm{\sigma})$ mimics as closely as
possible the data distribution in each basis, i.e. $|\psi_{\bm{\kappa}}(\bm{\sigma}^{[b]})|^{2}\simeq P_{b}(\bm{\sigma}^{[b]})$.
This is achieved by searching for the NN parameters that minimize
the total statistical divergence $\Xi(\bm{\kappa})$ between the target
distributions and the reconstructed ones. Several possible choices
can be made for $\Xi(\bm{\kappa})$. Here, we define it as the sum
of the Kullbach-Leibler (KL) divergences in each basis: 
\begin{equation}
\Xi(\bm{\kappa})\equiv\sum_{b=0}^{N_{B}}\mathbb{KL}_{\bm{\kappa}}^{[b]}=\sum_{b=0}^{N_{B}}\sum_{\{\bm{\sigma}^{[b]}\}}P_{b}(\bm{\sigma}^{[b]})\log\frac{P_{b}(\bm{\sigma}^{[b]})}{|\psi_{\bm{\kappa}}(\bm{\sigma}^{[b]})|^{2}}.\label{KL_full}
\end{equation}
The total divergence $\Xi(\bm{\kappa})$ is positive definite, and
attains the minimum value of $0$ when the reconstruction is perfect
in each basis: $|\psi_{\bm{\kappa}}(\bm{\sigma}^{[b]})|^{2}=P_{b}(\bm{\sigma}^{[b]})$.
Depending on the target wave-function, a sufficiently large set of
measurement bases must be included, in order to have enough information
to estimate the phases in the reference basis. In practice, for most
states of interest it is enough to include a number of bases which
scales only polynomially with system size.

Once the training is complete, the NN provides a compact representation
$\psi_{\bm{\kappa}}(\bm{\sigma})$ of the target wave-function $\Psi(\bm{\sigma})$.
In turn, this representation can be used to efficiently compute various
observables of interest, overlaps with other known quantum states
and virtually any other information not directly accessible in the
experiment. In the next two sub-sections we describe in details the
specific parametrization of the NN wave-function adopted in this work
and its optimization.

\subsection{The RBM wave-function}

\begin{figure}[t]
\noindent \centering{}\includegraphics[width=0.5\columnwidth]{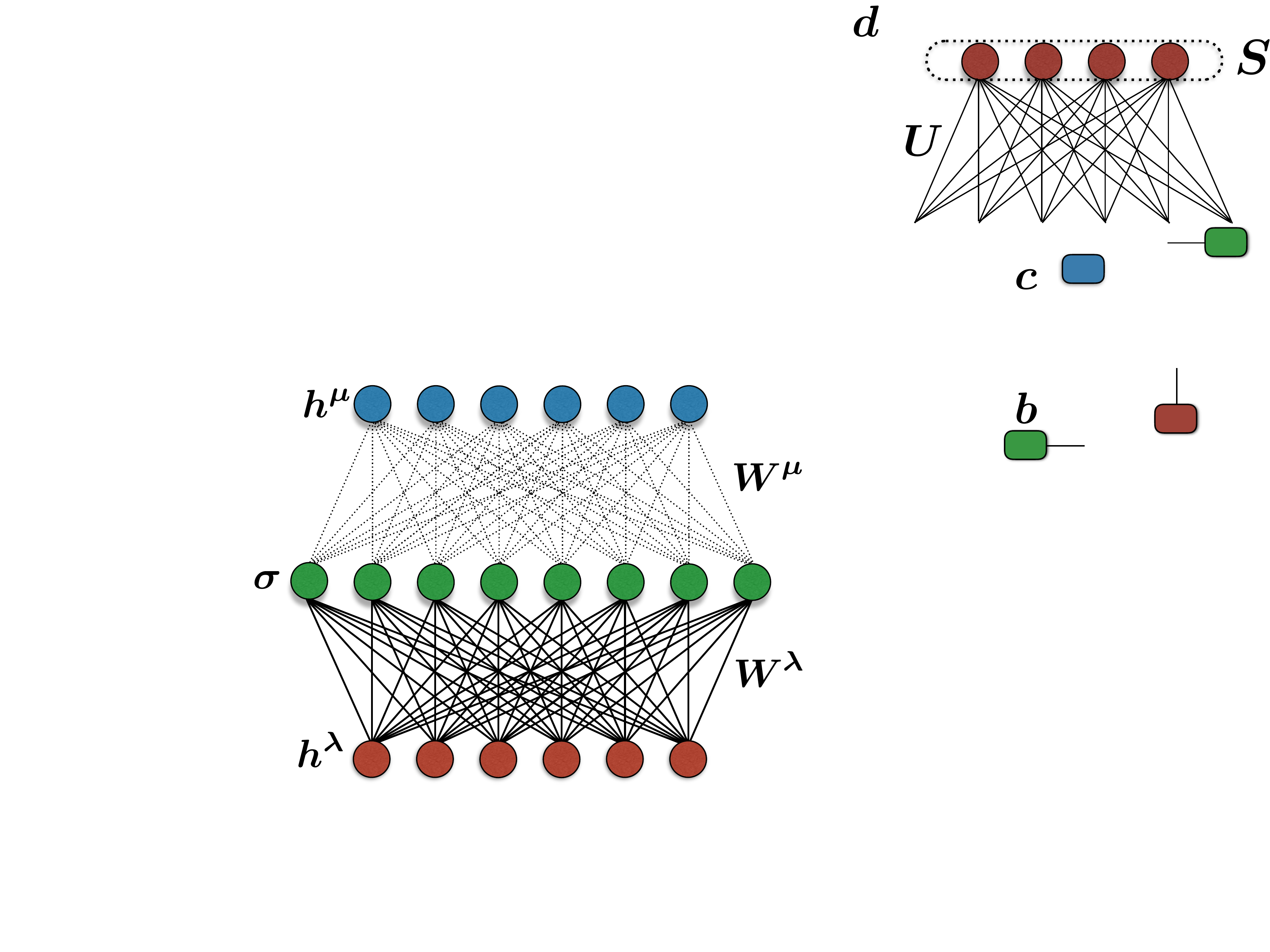}
\caption{RBM parametrization of the wave-function: a layer $\bm{\sigma}$ of
neurons describing the physical variables (e.g. spins, particles,
etc) is connected to two hidden layers $\bm{h^{\lambda}}$ and $\bm{h^{\mu}}$
with weights $\bm{W^{\lambda}}$ and $\bm{W^{\mu}}$ (external fields
are not drawn in the figure). Thick lines refer to the weighted connections
employed in the sampling of configurations $\bm{\sigma}$, while dotted
lines are used to parametrize the phase $\phi_{\mu}$. }
\label{RBM} 
\end{figure}

The are many possible architecture and NN that can be employed to
represent a quantum many-body state. We decide to employ a powerful
stochastic NN called a restricted Boltzmann machine (RBM). The network
architecture of a RBM features two layers of stochastic binary neurons,
a visible layer $\bm{\sigma}$ describing the physical variables and
a hidden layer $\bm{h}$. The expressive power of the model can be
characterized by the ratio $\alpha=M/N$ between the number of hidden
neurons $M$ and visible neurons $N$. A RBM is also an energy-based
model, sharing many properties of physical model in statistical mechanics.
In particular, it associates to the graph structure a probability
distribution given by the Boltzmann distribution 
\begin{equation}
p_{\bm{\kappa}}(\bm{\sigma},\bm{h})=\frac{1}{Z_{\bm{\kappa}}}\,\mbox{e}^{\,\sum_{ij}\,W_{ij}^{\kappa}h_{i}\sigma_{j}+\sum_{j}\,b_{j}^{\kappa}\sigma_{j}+\sum_{i}\,c_{i}^{\kappa}h_{i}},
\end{equation}
where $Z_{\bm{\kappa}}=\sum_{\bm{\sigma},\bm{h}}p_{\bm{\kappa}}(\bm{\sigma},\bm{h})$
is the normalization constant and $\bm{\kappa}$ now consists on the
weights $\bm{W}^{\kappa}$ connecting the two layers and the fields
(biases) $\bm{b}^{\kappa}$ and $\bm{c}^{\kappa}$ coupled to each
visible and hidden neurons, respectively. The distribution (of interest)
over the visible layer is obtained by marginalization over the hidden
degrees of freedom 
\begin{equation}
p_{\bm{\kappa}}(\bm{\sigma})=\sum_{\{\bm{h}\}}p_{\bm{\kappa}}(\bm{\sigma},\bm{h})=\mbox{e}^{\sum_{j}b_{j}^{\kappa}\sigma_{j}+\sum_{i}\log\left(1+\text{e}^{\,c_{i}^{\kappa}+\sum_{j}\,W_{ij}^{\kappa}\sigma_{j}}\right)}.
\end{equation}

The RBM wave-function is then defined as 
\begin{equation}
\psi_{\bm{\lambda},\bm{\mu}}(\bm{\sigma})=\sqrt{\frac{p_{\bm{\lambda}}(\bm{\sigma})}{Z_{\bm{\lambda}}}}\:\text{e}^{i\phi_{\bm{\mu}}(\bm{\sigma})/2},\label{psiNN}
\end{equation}
where $\phi_{\bm{\mu}}(\bm{\sigma})=\log p_{\bm{\mu}}(\bm{\sigma})$
and $\bm{\lambda},\bm{\mu}$ are the two set of parameters (Fig.~\ref{RBM}).
Note that the sampling of configurations $\bm{\sigma}$ from $|\psi_{\bm{\lambda},\bm{\mu}}(\bm{\sigma})|^{2}$,
involves only the amplitude distribution $p_{\bm{\lambda}}(\bm{\sigma})/Z_{\bm{\lambda}}$.
This can be achieved, as usual for RBMs, by performing Block Gibbs
sampling with the two conditional distributions $p_{\bm{\lambda}}(\bm{\sigma}\;|\;\bm{h})$
and $p_{\bm{\lambda}}(\bm{h}\;|\;\bm{\sigma})$, which can be computed
exactly. This procedure is very efficient since each neuron in one
layer of the RBM is connected only to neurons of a different layer,
thus enabling us to sample all units (in one layer) simultaneously.

\subsection{Gradients of the total divergence }

The first step in the RBMs trainings is to build the dataset of measurements.
In general, different basis are needed to estimate both amplitudes
and phases of the target state $\Psi(\bm{\sigma})$. We define a series
of datasets $D_{b}$ for each base $b=1,\dots,N_{B}$, with each dataset
$D_{b}=\{\bm{\sigma}_{i}^{[b]}\}_{i=1}^{|D_{b}|}$ consisting of $|D_{b}|$
density measurements with underlying distribution $P_{b}(\bm{\sigma}^{[b]})\propto|\Psi(\bm{\sigma}^{[b]})|^{2}$,
where $\bm{\sigma}^{[b]}=(\sigma_{1}^{[b]},\dots,\sigma_{N}^{[b]})$
and $\bm{\sigma}^{[0]}=\bm{\sigma}$. The quantity to minimize, also
called negative log-likelihood, is then 
\begin{equation}
\Xi(\bm{\kappa})=-\sum_{b=0}^{N_{B}}\frac{1}{|D_{b}|}\sum_{\bm{\sigma}^{[b]}\in D_{b}}\log|\psi_{\bm{\lambda},\bm{\mu}}(\bm{\sigma}^{[b]})|^{2}\label{KL_dataset}
\end{equation}
where we omitted here a constant term given by the sum of the the
cross-entropies of the datasets $\sum_{b}\mathbb{H}(D_{b})$. The
NN wave-function in the $\bm{\sigma}^{[b]}$ bases is simply obtained
by 
\begin{equation}
\psi_{\bm{\lambda},\bm{\mu}}(\bm{\sigma}^{[b]})=\sum_{\{\bm{\sigma}\}}U_{b}(\bm{\sigma},\bm{\sigma}^{[b]})\:\psi_{\bm{\lambda},\bm{\mu}}(\bm{\sigma}),
\end{equation}
with $U_{b}(\bm{\sigma},\bm{\sigma}^{[b]})$ being the basis transformation
matrix. The rotated state, $\psi_{\bm{\lambda},\bm{\mu}}(\bm{\sigma}^{[b]})$,
can be computed efficiently, provided that $U$ acts non-trivially
on a limited number of qubits.

We proceed now to give the expressions for the various gradients needed
in the training. By plugging Eq.~\ref{psiNN} in Eq.~\ref{KL_dataset}
we obtain 
\begin{equation}
\Xi(\bm{\lambda},\bm{\mu})=(N_{B}+1)\log Z_{\bm{\lambda}}-\sum_{b=0}^{N_{B}}\frac{1}{|D_{b}|}\sum_{\bm{\sigma}^{[b]}\in D_{b}}\left[\log\left(\sum_{\{\bm{\sigma}\}}U_{b}(\bm{\sigma},\bm{\sigma}^{[b]})\:\sqrt{p_{\bm{\lambda}}(\bm{\sigma})}\:\text{e}^{i\phi_{\bm{\mu}}(\bm{\sigma})/2}\right)+c.c.\right].
\end{equation}
We define now the two gradients 
\begin{eqnarray}
\mathcal{D}_{\bm{\lambda}}(\bm{\sigma}) & = & \frac{1}{p_{\bm{\lambda}}(\bm{\sigma})}\nabla_{\bm{\lambda}}p_{\bm{\lambda}}(\bm{\sigma})\\
\mathcal{D}_{\bm{\mu}}(\bm{\sigma}) & = & \nabla_{\bm{\mu}}\:\phi_{\bm{\mu}}(\bm{\sigma}),
\end{eqnarray}
where the derivatives of the RBM distribution are: 
\begin{equation}
\frac{\partial}{\partial W_{ij}}p_{\bm{\kappa}}(\bm{\sigma})=\frac{\sigma_{j}}{1+\text{e}^{-\sum_{j}W_{ij}^{\kappa}\sigma_{j}-c_{i}}}\:,
\end{equation}
\begin{equation}
\frac{\partial}{\partial b_{j}}p_{\bm{\kappa}}(\bm{\sigma})=\sigma_{j},
\end{equation}
and 
\begin{equation}
\frac{\partial}{\partial c_{i}}p_{\bm{\kappa}}(\bm{\sigma})=\frac{1}{1+\text{e}^{-\sum_{j}W_{ij}^{\kappa}\sigma_{j}-c_{i}}}\:.
\end{equation}
We also define the quasi-probability distribution 
\begin{equation}
Q_{b}(\bm{\sigma},\bm{\sigma}^{[b]})=U_{b}(\bm{\sigma},\bm{\sigma}^{[b]})\:\sqrt{p_{\bm{\lambda}}(\bm{\sigma})}\:\text{e}^{i\phi_{\bm{\mu}}(\bm{\sigma})/2}\:.
\end{equation}
Then, the derivatives of the KL divergence with respect to the parameters
$\bm{\lambda}$ and $\bm{\mu}$ are 
\begin{equation}
\nabla_{\bm{\lambda}}\:\Xi(\bm{\lambda},\bm{\mu})=(N_{B}+1)\langle\mathcal{D}_{\bm{\lambda}}\rangle_{p_{\bm{\lambda}}}-\sum_{b=0}^{N_{B}}\frac{1}{|D_{b}|}\sum_{\bm{\sigma}^{[b]}\in D_{b}}\text{Re}\left\{ \langle\mathcal{D}_{\bm{\lambda}}\rangle_{Q_{b}}\right\} ,
\end{equation}
and 
\begin{equation}
\nabla_{\bm{\mu}}\:\Xi(\bm{\lambda},\bm{\mu})=\sum_{b=0}^{N_{B}}\frac{1}{|D_{b}|}\sum_{\bm{\sigma}^{[b]}\in D_{b}}\text{Im}\left\{ \langle\mathcal{D}_{\bm{\mu}}\rangle_{Q_{b}}\right\} .\label{eq:gradmu}
\end{equation}
In the expression above we have defined the pseudo-averages: 
\begin{eqnarray}
\langle\mathcal{D}_{\bm{\lambda/\mu}}\rangle_{Q_{b}} & = & \frac{\sum_{\{\bm{\sigma}\}}\mathcal{D}_{\bm{\lambda/}\bm{\mu}}(\bm{\sigma})Q_{b}(\bm{\sigma},\bm{\sigma}^{[b]})\psi_{\bm{\lambda},\bm{\mu}}(\bm{\bm{\sigma}})}{\sum_{\{\bm{\sigma}\}}Q_{b}(\bm{\sigma},\bm{\sigma}^{[b]})\psi_{\bm{\lambda},\bm{\mu}}(\bm{\bm{\sigma}})},
\end{eqnarray}
which can be efficiently computed directly summing over the samples
in the datasets $D_{b}$. On the other hand, the evaluation of the
average 
\begin{equation}
\langle\mathcal{D}_{\bm{\lambda}}\rangle_{p_{\bm{\lambda}}}=\frac{1}{Z_{\bm{\lambda}}}\sum_{\{\bm{\sigma}\}}p_{\bm{\lambda}}(\bm{\sigma})\mathcal{D}_{\bm{\lambda}}(\bm{\sigma}),
\end{equation}
requires the knowledge of the normalization constant $Z_{\bm{\lambda}}$,
which is not directly accessible. However, as per standard RBM training,\cite{Hinton02}
one can approximate this average by 
\begin{equation}
\langle\mathcal{D}_{\bm{\lambda}}\rangle_{p_{\bm{\lambda}}}\simeq\frac{1}{n}\sum_{k=1}^{n}\:\mathcal{D}_{\bm{\lambda}}(\bm{\sigma}_{k}),\label{eq:derlambda}
\end{equation}
where $\bm{\sigma}_{k}$ are samples generated using a Markov-chain
Monte Carlo simulation.

Finally, we point out that in our work we have adopted a slightly
simplified training scheme. In particular, we break down the training
into two steps. First, we learn the amplitudes only by optimizing
the parameters $\bm{\lambda}$. In this case, it is sufficient to
minimize the KL divergence over the reference basis only (i.e. $\bm{\sigma}$).
This part of the training is to all purposes a standard unsupervised
learning procedure, involving the generation of samples from the RBM.\cite{Goodfellow-et-al-2016}
Then, we fix the parameters $\bm{\lambda}$, and use the measurements
in the auxiliary bases to determine the optimal values of the phase
parameters $\bm{\mu}$. This other part of the training is achieved
using the gradient in Eq. \ref{eq:gradmu}, thus not requiring Monte
Carlo sampling from the NN.

\subsection{Training the neural network}

For a given set of parameters (e.g. $\bm{\mu}$), the easiest way
to numerically minimize the total divergence, Eq. \ref{KL_dataset},
is by using simple stochastic gradient descent\cite{Goodfellow-et-al-2016}.
Each parameter $\mu_{j}$ is updated as 
\begin{equation}
\mu_{j}\leftarrow\mu_{j}-\eta\:\langle g_{j}\rangle_{B},
\end{equation}
where the gradient step $\eta$ is called learning rate and the gradient
$g_{j}$ is averaged over a batch $B$ ($|B|\ll|D|$) of samples drawn
randomly from the full dataset: 
\begin{equation}
\langle g_{j}\rangle_{B}=\frac{1}{|B|}\sum_{\bm{\sigma}\in B}\text{Im}\left\{ \langle\mathcal{D}_{\mu_{j}}\rangle_{Q_{b}}\right\} .
\end{equation}
Stochastic gradient descent was the optimization method used to learn
the amplitudes of each physical system presented in the paper. For
the learning of the phases however, we instead implemented the natural
gradient descent \cite{amari_natural_1998}, which revealed to be
more effective, though at the cost of increased computational resources.
In this case we update the parameters as 
\begin{equation}
\mu_{j}\leftarrow\mu_{j}-\eta\:\sum_{i}\langle S_{ij}^{-1}\rangle_{B}\:\langle g_{j}\rangle_{B},
\end{equation}
where we have introduced the Fisher information matrix: 
\begin{equation}
\langle S_{ij}\rangle_{B}=\frac{1}{|B|}\sum_{\bm{\sigma}\in B}\text{Im}\left\{ \langle\mathcal{D}_{\mu_{i}}\rangle_{Q_{b}}\right\} \text{Im}\left\{ \langle\mathcal{D}_{\mu_{j}}\rangle_{Q_{b}}\right\} .
\end{equation}
The learning rate magnitude $\eta$ is set to 
\begin{equation}
\eta=\frac{\eta_{0}}{\sqrt{\sum_{ij}\langle S_{ij}\rangle_{B}\times\langle g_{i}\rangle_{B}\langle g_{j}\rangle_{B}}}
\end{equation}
with some initial learning rate $\eta_{0}$. The matrix $\langle S_{ij}\rangle_{B}$
takes into account the fact that, since the parametric dependence
of the RBM function is non-linear, a small change of some parameters
may correspond to a very large change of the distribution. In this
way one implicitly uses an adaptive learning rate for each parameter
$\mu_{j}$ and speed-up the optimization compared to the simplest
gradient descent. We notice that a very similar technique is successfully
used in Quantum Monte Carlo for optimizing high-dimensional variational
wave-functions\cite{sorella1998green}. Similarly to our case, noisy
gradients, which come from the Monte Carlo statistical evaluation
of energy derivatives with respect to the parameters, are present,
while the matrix $S$ is instead given by the covariance matrix of
these forces. Since the matrix $\langle S_{ij}\rangle_{B}$ is affected
by statistical noise, we regularize it by adding a small diagonal
offset, thus improving the stability of the optimization.

\subsection{Training datasets}

In our work we have benchmarked NN-QST on artificial datasets, consisting
of a collection of independent measurements obtained by projecting
the physical system wave-function $|\Psi\rangle$ into the various
basis $\{\bm{\sigma}^{[b]}\}$. Whenever possible, we perform exact
sampling of the full wave-function $\Psi(\bm{\sigma})$, that is when
the system size is small enough, or the wave-function itself is simple
enough (e.g. W state). In the case of QST for ground states of local
Hamiltonian, we investigated system sizes out of the reach of any
exact diagonalization techniques, and we therefore build the datasets
using quantum Monte Carlo (QMC) simulations.

We use the Path-Integral Monte Carlo (PIMC) variant of the QMC family
of algorithms, which allows us to sample from the exact ground state
density distribution $|\Psi(\bm{\sigma})|^{2}$ for the Transverse
field Ising model (TFIM) and the anisotropic Heisenberg model (XXZ),
whose Hamiltonians $\mathcal{H}$ are defined in the main text. The
PIMC method relies on the property that the partition functions of
these $d$-dimensional quantum spin-$\frac{1}{2}$ systems can be
mapped onto that of $(d+1)$-dimensional classical systems.\cite{suzuki76}
The additional dimension is called \textquotedbl{}imaginary time\textquotedbl{}
$\tau$, which goes from $0$ to $\beta=1/T$, i.e. the inverse physical
temperature of the model. In this work we employ the discrete-time
version of the PIMC algorithm, where the total \textquotedbl{}imaginary
time\textquotedbl{} $\beta$ is discretized in $M_{\tau}$ steps,
and the simulations are exact in the $\beta/M_{\tau}\rightarrow0$
limit. Therefore the quantum simulations of the $N$ spins TFIM is
mapped onto a system of $N\times M_{\tau}$ classical spin variables,
with suitable interactions along the \textquotedbl{}imaginary time\textquotedbl{}
direction (see Ref. \onlinecite{suzuki76} for details).

Classical Metropolis Monte Carlo (MC) on this larger system can then
be performed in order to collect samples of the quantum distribution
in the $\{\bm{\sigma}\}$ basis. Since we are interested in the ground
state distribution, we use a sufficiently large inverse temperature,
in the range $\beta=10-20$ and a converged number of $M_{\tau}=1024-2048$.
Statistically independent samples are collected during each MC simulation
waiting for a sufficiently large number of MC moves, i.e. larger than
the autocorrelation time of the Markov chain. In order to decrease
the autocorrelations between successive MC configurations we use cluster
update algorithms. In the case of the TFIM we use the Wolff single
cluster algorithm\cite{PhysRevLett.62.361}. Here clusters can be
un-restricted in the volume or restricted in such a way to extend
only along the \textquotedbl{}imaginary time\textquotedbl{} direction\cite{Rieger1999,mazzola2017quantum}.
Both choices drastically improve the efficiency compared to the simple
local update scheme. For the XXZ model we use a single cluster update
version of the Loop algorithm\cite{evertz_loop_2003}.

\section{Cases of Study}

We now describe the details concerning trainings and the measurements
for the physical systems investigated in the main paper.

\subsection{W state}

The $N$-qubits W state 
\begin{equation}
|\Psi_{W}\rangle=\frac{1}{\sqrt{N}}\bigg(|100\dots\rangle+|010\dots\rangle+\dots+|0\dots01\rangle\bigg),
\end{equation}
can be efficiently sampled to generate the training datasets, irrespectively
on the system size $N$. Moreover, since each coefficient $\Psi_{W}(\bm{\sigma})$
is real and positive, we only need to learn the amplitudes and we
can adopt the reduced a simpler version of the RBM wave-function,
that is 
\begin{equation}
\psi_{\bm{\lambda}}(\bm{\sigma})=\sqrt{\frac{p_{\bm{\lambda}}(\bm{\sigma})}{Z_{\bm{\lambda}}}},
\end{equation}
thus using only one set ($\bm{\lambda}$) of network parameters.

To quantify the performances of the training we compute the overlap
$O$ between the W state wave-function and the RBM wave-function 
\begin{equation}
O=\langle\Psi_{W}|\psi_{\bm{\lambda}}\rangle=\sum_{\bm{\sigma}}\:\Psi_{W}(\bm{\sigma})\;\psi_{\bm{\lambda}}(\bm{\sigma}),\label{fullOverlap}
\end{equation}
where $\Psi_{W}({\bm{\sigma}})=\delta(\bm{\sigma}-2^{k})/\sqrt{N}$
for $k\in(0,\dots,N-1)$. As we cannot perform the full sum in Eq.~\ref{fullOverlap}
for large system sizes $N$, and we do not know the normalization
constant $Z_{\bm{\lambda}}$, we instead compute the square of the
overlap as 
\begin{equation}
\begin{split}O^{2} & =\frac{\langle\Psi_{W}|\psi_{\bm{\lambda}}\rangle}{\langle\psi_{\bm{\lambda}}|\psi_{\bm{\lambda}}\rangle}\times\frac{\langle\Psi_{W}|\psi_{\bm{\lambda}}\rangle}{\langle\Psi_{W}|\Psi_{W}\rangle}\\
 & =\frac{\sum_{\bm{\sigma}}\;|\psi_{\bm{\lambda}}(\bm{\sigma})|^{2}\;\frac{\Psi_{W}(\bm{\sigma})}{\psi_{\bm{\lambda}}(\bm{\sigma})}}{\sum_{\bm{\sigma}}\;|\psi_{\bm{\lambda}}(\bm{\sigma})|^{2}}\times\frac{\sum_{\bm{\sigma}}\;|\Psi_{W}(\bm{\sigma})|^{2}\;\frac{\psi_{\bm{\lambda}}(\bm{\sigma})}{\Psi_{W}(\bm{\sigma})}}{\sum_{\bm{\sigma}}\;|\Psi_{W}(\bm{\sigma})|^{2}}\\
 & =\left\langle \frac{\Psi_{W}(\bm{\sigma})}{\sqrt{p_{\bm{\lambda}}(\bm{\sigma})}}\right\rangle {}_{p_{\bm{\lambda}}}\times\left\langle \frac{\sqrt{p_{\bm{\lambda}}(\bm{\sigma})}}{\Psi_{W}(\bm{\sigma})}\right\rangle {}_{|\Psi_{W}|^{2}}\\
 & =\left(\frac{1}{n}\sum_{j=1}^{n}\frac{1}{\sqrt{p_{\bm{\lambda}}(\bm{\sigma}_{j})}}\sum_{k=0}^{N-1}\frac{\delta(\bm{\sigma}_{j}-2^{k})}{\sqrt{N}}\right)\times\left(\sum_{k=0}^{N-1}\sqrt{\frac{p_{\bm{\lambda}}(\bm{\sigma}=2^{k})}{N}}\right),
\end{split}
\end{equation}
where the qubits configurations $\bm{\sigma}_{j}$ are drawn directly
from the trained RBM distribution $p_{\bm{\lambda}}(\bm{\sigma})$
by performing block Gibbs sampling from the two conditional distributions
$p_{\bm{\lambda}}(\bm{\sigma}\;|\;\bm{h})$ and $p_{\bm{\lambda}}(\bm{h}\;|\;\bm{\sigma})$.

We now consider the case where local phase shifts with random phases
$\theta_{j}$ are applied to the W state: 
\begin{equation}
|\tilde{\Psi}_{W}\rangle=\frac{1}{\sqrt{N}}\bigg(\text{e}^{i\theta_{1}}|100\dots\rangle+\text{e}^{i\theta_{2}}|010\dots\rangle+\dots+\text{e}^{i\theta_{N}}|0\dots01\rangle\bigg).
\end{equation}
In this case, we use the full wave-function $\psi_{\bm{\lambda},\bm{\mu}}(\bm{\sigma})$
to learn both amplitude and phases. Given the structure of the state
$|\tilde{\Psi}_{W}\rangle$, we require the $(N-1)$ supplementary
basis 
\begin{equation}
\{X,X,Z,Z,\dots\}\,,\,\{Z,X,X,Z,\dots\}\,,\,\{Z,Z,X,X,\dots\},
\end{equation}
where in the basis $\{X_{j},X_{j+1}\}$ we have $|\tilde{\Psi}_{W}|^{2}\propto\cos(\theta_{j+1}-\theta_{j})$,
and the $(N-1)$ supplementary basis 
\begin{equation}
\{X,Y,Z,Z,\dots\}\,,\,\{Z,X,Y,Z,\dots\}\,,\,\{Z,Z,X,Y,\dots\},
\end{equation}
where in the basis $\{X_{j},Y_{j+1}\}$ we have $|\tilde{\Psi}_{W}|^{2}\propto\sin(\theta_{j+1}-\theta_{j})$.
The RBM is then trained on a total of $2N-1$ basis (including the
standard basis for the amplitude learning). The transformation matrices
for the $j$-th basis ($\{X_{j},X_{j+1}\}$) and ($\{X_{j},Y_{j+1}\}$)
are given by $U_{j}^{XX}=H_{j}\otimes H_{j+1}$ and $U_{j}^{XY}=H_{j}\otimes K_{j+1}$
respectively, where 
\begin{equation}
H=\frac{1}{\sqrt{2}}\begin{bmatrix}1 & 1\\
1 & -1
\end{bmatrix}\qquad\qquad K=\frac{1}{\sqrt{2}}\begin{bmatrix}1 & -i\\
1 & i
\end{bmatrix},
\end{equation}

The average of the gradient $\mathcal{D}_{\bm{\mu}}$ over the quasi-probability
distribution $Q_{j}^{X\nu}$ ($\nu=X,Y$) is then given by 
\begin{equation}
\begin{split}\langle\mathcal{D}_{\bm{\mu}}\rangle_{Q_{j}^{X\nu}} & =\frac{\sum_{\bm{\sigma}}U_{j}^{X\nu}(\bm{\sigma},\bm{\sigma}^{[j]})\mathcal{D}_{\mu}(\bm{\sigma})\psi_{\bm{\lambda},\bm{\mu}}(\bm{\sigma})}{\sum_{\bm{\sigma}}U_{j}^{X\nu}(\bm{\sigma},\bm{\sigma}^{[j]})\psi_{\bm{\lambda},\bm{\mu}}(\bm{\sigma})}\\
 & =\frac{1}{\Lambda^{j}(\bm{\sigma}^{[j]})}\bigg[\mathcal{D}_{\bm{\mu}}(\bm{\sigma}_{00}^{[j]})\xi_{\bm{\lambda},\bm{\mu}}(\bm{\sigma}_{00}^{[j]})+i^{\delta_{\nu,X}}(1-2\sigma_{j+1}^{\nu})\mathcal{D}_{\bm{\mu}}(\bm{\sigma}_{01}^{[j]})\xi_{\bm{\lambda},\bm{\mu}}(\bm{\sigma}_{01}^{[j]})+\\
 & +(1-2\sigma_{j}^{x})\mathcal{D}_{\bm{\mu}}(\bm{\sigma}_{10}^{[j]})\xi_{\bm{\lambda},\bm{\mu}}(\bm{\sigma}_{10}^{[j]})+i^{\delta_{\nu,X}}(1-2\sigma_{j+1}^{\nu})(1-2\sigma_{j}^{x})\mathcal{D}_{\bm{\mu}}(\bm{\sigma}_{11}^{[j]})\xi_{\bm{\lambda},\bm{\mu}}(\bm{\sigma}_{11}^{[j]})\bigg]\\
\end{split}
\end{equation}
where we defined 
\begin{equation}
\begin{split}\Lambda^{j}(\bm{\sigma}^{[j]}) & =\xi_{\bm{\lambda},\bm{\mu}}(\bm{\sigma}_{00}^{[j]})+i^{\delta_{\nu,X}}(1-2\sigma_{j+1}^{\nu})(1-2\sigma_{j}^{x})\xi_{\bm{\lambda},\bm{\mu}}(\bm{\sigma}_{11}^{[j]})+\\
 & +i^{\delta_{\nu,X}}(1-2\sigma_{j+1}^{\nu})\xi_{\bm{\lambda},\bm{\mu}}(\bm{\sigma}_{01}^{[j]})+(1-2\sigma_{j}^{x})\xi_{\bm{\lambda},\bm{\mu}}(\bm{\sigma}_{10}^{[j]})\:,
\end{split}
\end{equation}
\begin{equation}
\xi_{\bm{\lambda},\bm{\mu}}(\bm{\sigma}_{\alpha\beta}^{[j]})=\bigg(\sum_{\alpha=0,1}\sum_{\beta=0,1}\;\sqrt{p_{\bm{\lambda}}(\bm{\sigma}_{\alpha\beta}^{[j]})}\bigg)^{-1}\sqrt{p_{\bm{\lambda}}(\bm{\sigma}_{\alpha\beta}^{[j]})}\:\text{e}^{i\phi_{\bm{\mu}}(\bm{\sigma}_{\alpha\beta}^{[j]})/2},
\end{equation}
and $\bm{\sigma}_{\alpha\beta}^{[j]}=(\sigma_{1}^{z},\dots,\sigma_{j}^{z}=\alpha,\sigma_{j+1}^{z}=\beta,\dots\sigma_{N}^{z}$).

\subsection{Magnetic observables of local Hamiltonians}

The many-body Hamiltonians considered in this work are the TFIM and
the XXZ model. In both cases, the ground state wave-function is real
and positive, which means we can again restrict ourselves to learn
the amplitudes with one set of parameters, i.e. using the RBM wave-function
$\psi_{\bm{\lambda}}(\bm{\sigma})$. Instead of computing the overlap,
which is clearly intractable for the system sizes of interest, we
evaluate the quality of the QST by comparing different magnetic observables
computed using the RBM wave-function, with results obtained with Quantum
Monte Carlo (QMC) simulations.

Given some observable $\mathcal{O}=\sum_{\bm{\sigma},\bm{\sigma}^{\prime}}\mathcal{O}_{\bm{\sigma\sigma}^{\prime}}|\bm{\sigma}\rangle\langle\bm{\sigma}^{\prime}|$,
we can calculate its expectation value using the RBM wave-function
as 
\begin{equation}
\langle\mathcal{O}\rangle=\sum_{\bm{\sigma},\bm{\sigma}^{\prime}}\psi_{\bm{\lambda}}(\bm{\sigma})\psi_{\bm{\lambda}}(\bm{\sigma}^{\prime})\mathcal{O}_{\bm{\sigma\sigma}^{\prime}}.
\end{equation}
If the operator $\mathcal{O}$ is diagonal in the $\{\bm{\sigma}\}$
basis, i.e. $\mathcal{O}_{\bm{\sigma\sigma}^{\prime}}^{D}=\mathcal{O}(\bm{\sigma})\delta_{\bm{\sigma\sigma}^{\prime}}$,
then 
\begin{equation}
\begin{split}\langle\mathcal{O}^{D}\rangle & =\sum_{\bm{\sigma}}|\psi_{\bm{\lambda}}(\bm{\sigma})|^{2}\mathcal{O}(\bm{\sigma})=\frac{1}{Z_{\bm{\lambda}}}\sum_{\bm{\sigma}}p_{\bm{\lambda}}(\bm{\sigma})\mathcal{O}(\bm{\sigma})\\
 & \simeq\frac{1}{n}\sum_{k=1}^{n}\mathcal{O}(\bm{\sigma}_{k}),
\end{split}
\end{equation}
where $\bm{\sigma}_{k}$ are sampled directly with the RBM. If, on
the other hand, the operator $\mathcal{O}$ is off-diagonal, we can
still compute its expectation value, provided that its matrix representation
in the $\{\bm{\sigma}\}$ basis is sparse. In this case, we obtain
\begin{equation}
\langle\mathcal{O}^{ND}\rangle=\sum_{\bm{\sigma}}|\psi_{\bm{\lambda}}(\bm{\sigma})|^{2}\mathcal{O}_{L}(\bm{\sigma})\simeq\frac{1}{n}\sum_{k=1}^{n}\mathcal{O}_{L}(\bm{\sigma}_{k}),
\end{equation}
where 
\begin{equation}
\mathcal{O}_{L}(\bm{\sigma})=\sum_{\bm{\sigma}^{\prime}}\sqrt{\frac{p_{\bm{\lambda}}(\bm{\sigma}^{\prime})}{p_{\bm{\lambda}}(\bm{\sigma})}}\mathcal{O}_{\bm{\sigma\sigma}^{\prime}},
\end{equation}
is the so-called \textquotedbl{}local estimate\textquotedbl{} of $\mathcal{O}$.
For the TFIM we compare the value of the off-diagonal transverse field
magnetization $\langle\sigma^{x}\rangle=\sum_{i=1}^{N}\langle\sigma_{i}^{x}\rangle$,
with its QMC estimate obtained following the path-integral formulation
of the expectation value of non-diagonal operators (see Ref. \onlinecite{PhysRevB.78.134428}
for its explicit derivation).

\subsection{Unitary evolution}

In the previous section we discussed QST of ground state wave-functions
for many-body Hamiltonians. In addition to this case, we have also
investigated the unitary dynamics induced by Hamiltonian evolution.
We consider the quantum ``quench'' setting, where the physical system
is prepared in a state $|\Psi_{0}\rangle$ and it is time evolved
with an Hamiltonian $\mathcal{H}$, leading to the state: 
\begin{equation}
|\Psi(t)\rangle=\text{e}^{-i\mathcal{H}t}|\Psi_{0}\rangle.
\end{equation}
In this case, for some fixed time $t$ we build a dataset of spins
density measurements $P_{b}(\bm{\sigma},t)=|\Psi(\bm{\sigma}^{[b]},t)|^{2}$
and train the RBM to learn $\Psi(\bm{\sigma},t)$. Since, because
of the time evolution operator, the state is complex-valued, we once
again employ the full RBM wave-function $\psi_{\bm{\lambda},\bm{\mu}}(\bm{\sigma})$
for the QST. In this case, the bases $\{Z,\dots,X_{j},\dots,Z\}$
and $\{Z,\dots,Y_{j},\dots,Z\}$, respectively with bases rotations
$U_{j}^{X}=H_{j}$ and $U_{j}^{Y}=K_{j}$, are sufficient to reconstruct
the wave-function phases.

The average of the gradient $\mathcal{D}_{\bm{\mu}}$ over the quasi-probability
distribution $Q_{j}^{\nu}$ ($\nu=X,Y$) is now given by: 
\begin{equation}
\begin{split}\langle\mathcal{D}_{\bm{\mu}}\rangle_{Q_{j}^{\nu}} & =\frac{\sum_{\bm{\sigma}}U_{j}^{\nu}(\bm{\sigma},\bm{\sigma}^{[j]})\mathcal{D}_{\mu}(\bm{\sigma})\psi_{\bm{\lambda},\bm{\mu}}(\bm{\sigma})}{\sum_{\bm{\sigma}}U_{j}^{\nu}(\bm{\sigma},\bm{\sigma}^{[j]})\psi_{\bm{\lambda},\bm{\mu}}(\bm{\sigma})}\\
 & =\frac{\mathcal{D}_{\bm{\mu}}(\bm{\sigma}_{0}^{[j]})+i^{\delta_{\nu,X}}(1-2\sigma_{j}^{\nu})\mathcal{D}_{\bm{\mu}}(\bm{\sigma}_{1}^{[j]})\xi_{\bm{\lambda},\bm{\mu}}(\bm{\sigma}^{[j]})}{1+i^{\delta_{\nu,X}}(1-2\sigma_{j}^{\nu})\xi_{\bm{\lambda},\bm{\mu}}(\bm{\sigma}^{[j]})},
\end{split}
\end{equation}
where we have defined 
\begin{equation}
\xi_{\bm{\lambda},\bm{\mu}}(\bm{\sigma}^{[j]})=\sqrt{\frac{p_{\bm{\lambda}}(\bm{\sigma}_{1}^{[j]})}{p_{\bm{\lambda}}(\bm{\sigma}_{0}^{[j]})}}\:\text{e}^{i(\phi_{\bm{\mu}}(\bm{\sigma}_{1}^{[j]})-\phi_{\bm{\mu}}(\bm{\sigma}_{0}^{[j]})/2},
\end{equation}
and $\bm{\sigma}_{\alpha}^{[j]}=(\sigma_{1}^{z},\dots,\sigma_{j}^{z}=\alpha,\dots\sigma_{N}^{z}$).

\subsection{Entanglement entropy}

Given a bipartition of the physical system in a subregion $A$ and
its complement $A^{\perp}$, the (generalized) entanglement Renyi
entropies are defined as 
\begin{equation}
S_{\alpha}(\rho_{A})=\frac{1}{1-\alpha}\log[\text{Tr}(\rho_{A}^{\alpha})],
\end{equation}
where $\rho_{A}=\text{Tr}_{A^{\perp}}(\rho)$ is the reduced density
matrix for subregion A. We can write the RBM wave-function (previously
trained in the usual way) in the general from 
\begin{equation}
|\psi_{\bm{\lambda}}\rangle=\sum_{\bm{\alpha},\bm{\alpha}^{\perp}}\Gamma_{\bm{\alpha\alpha}^{\perp}}^{\bm{\lambda}}|\bm{\alpha}\rangle\otimes|\bm{\alpha}^{\perp}\rangle,
\end{equation}
where $\{\bm{\alpha}\}$ and $\{\bm{\alpha}^{\perp}\}$ are basis
state for the subregions $A$ and $A^{\perp}$ respectively.

For simplicity, we also assume in the following that the wave-function
$\Psi$ is real and positive. We then consider two non-interacting
copies of the physical system in a product state and introduce the
Swap operator 
\begin{equation}
\begin{split}\text{Swap}_{A}\,|\psi_{\bm{\lambda}}\rangle\otimes|\psi_{\bm{\lambda}}\rangle & =\text{Swap}_{A}\bigg(\sum_{\bm{\alpha}_{1},\bm{\alpha}_{1}^{\perp}}\Gamma_{\bm{\alpha}_{1}\bm{\alpha}_{1}^{\perp}}^{\bm{\lambda}}|\bm{\alpha}_{1}\rangle\otimes|\bm{\alpha}_{1}^{\perp}\rangle\bigg)\otimes\bigg(\sum_{\bm{\alpha}_{2},\bm{\alpha}_{2}^{\perp}}\Gamma_{\bm{\alpha}_{2}\bm{\alpha}_{2}^{\perp}}^{\bm{\lambda}}|\bm{\alpha}_{2}\rangle\otimes|\bm{\alpha}_{2}^{\perp}\rangle\bigg)\\
 & =\sum_{\bm{\alpha}_{1},\bm{\alpha}_{1}^{\perp}}\sum_{\bm{\alpha}_{2},\bm{\alpha}_{2}^{\perp}}\Gamma_{\bm{\alpha}_{1}\bm{\alpha}_{1}^{\perp}}^{\bm{\lambda}}\Gamma_{\bm{\alpha}_{2}\bm{\alpha}_{2}^{\perp}}^{\bm{\lambda}}(|\bm{\alpha}_{1}\rangle\otimes|\bm{\alpha}_{2}^{\perp})\otimes(|\bm{\alpha}_{2}\rangle\otimes|\bm{\alpha}_{1}^{\perp}).
\end{split}
\end{equation}
It follows that the expectation value of the Swap operator is 
\begin{equation}
\begin{split}\langle\text{Swap}_{A}\rangle & =\sum_{\bm{\alpha}_{1},\bm{\alpha}_{1}^{\perp}}\sum_{\bm{\alpha}_{2},\bm{\alpha}_{2}^{\perp}}\Gamma_{\bm{\alpha}_{1}\bm{\alpha}_{1}^{\perp}}^{\bm{\lambda}}(\Gamma_{\bm{\alpha}_{2}\bm{\alpha}_{2}^{\perp}}^{\bm{\lambda}})^{*}\Gamma_{\bm{\alpha}_{2}\bm{\alpha}_{2}^{\perp}}^{\bm{\lambda}}(\Gamma_{\bm{\alpha}_{2}\bm{\alpha}_{2}^{\perp}}^{\bm{\lambda}})^{*}\\
 & =\text{Tr}(\rho_{A}^{2})=\text{e}^{-S_{2}(\rho_{A})}.
\end{split}
\end{equation}
\begin{figure}[t]
\noindent \centering{}\includegraphics[width=0.75\columnwidth]{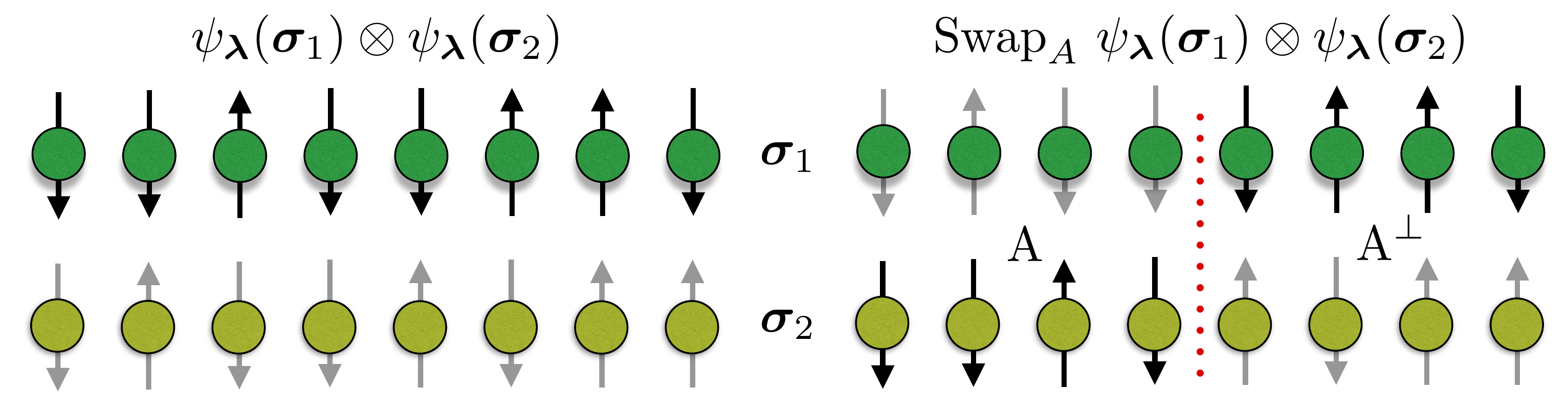}
\caption{Replica trick for the evaluation of the second Renyi entropy. }
\label{entrop} 
\end{figure}

Although the replica trick shown above already provides a way to compute
the entanglement entropy, the expectation values of the Swap operator
becomes very small when the subregion size grows larger, leading to
very high sampling noise. To avoid this issue, we implement the improved
ratio trick, proposed in Ref. \onlinecite{hastings_measuring_2010}.
Assuming we are dealing with a 1d chain with N sites, the entanglement
entropy for a subregion A of $n$ sites can be computed as 
\begin{equation}
S_{2}(\rho_{A})=-\sum_{j=0}^{n-1}\log\frac{\langle\text{Swap}_{A^{j+1}}\rangle}{\langle\text{Swap}_{A^{j}}\rangle},
\end{equation}
where $A^{j}$ contains $j$ sites and $\langle\text{Swap}_{A^{0}}\rangle=1$.
To estimate such expectation values we consider again the expansion
of the composite wave-function 
\begin{equation}
|\psi_{\bm{\lambda}}\rangle\otimes|\psi_{\bm{\lambda}}\rangle=\sum_{\bm{\sigma}_{1}}\sum_{\bm{\sigma}_{2}}\psi_{\bm{\lambda}}(\bm{\sigma}_{1})\psi_{\bm{\lambda}}(\bm{\sigma}_{2})|\bm{\sigma}_{1}\rangle\otimes|\bm{\sigma}_{2}\rangle,
\end{equation}
on which the expectation value of the Swap operator is 
\begin{equation}
\langle\text{Swap}_{A}^{j}\rangle=\sum_{\bm{\sigma}_{1}}\sum_{\bm{\sigma}_{2}}\psi_{\bm{\lambda}}(\bm{\sigma}_{1})\psi_{\bm{\lambda}}(\bm{\sigma}_{2})\psi_{\bm{\lambda}}(\bm{\sigma}_{12}^{j})\psi_{\bm{\lambda}}(\bm{\sigma}_{21}^{j}),
\end{equation}
where we defined $\bm{\sigma}_{12}^{j}=(\sigma_{1}^{1},\sigma_{1}^{2},\dots,\sigma_{1}^{j-1},\sigma_{2}^{j},\dots,\sigma_{2}^{N})$
and $\bm{\sigma}_{21}^{j}=(\sigma_{2}^{1},\sigma_{2}^{2},\dots,\sigma_{2}^{j-1},\sigma_{1}^{j},\dots,\sigma_{1}^{N})$.
The ratio of expectation values then can be rewritten as 
\begin{equation}
\begin{split}\frac{\langle\text{Swap}_{A}^{j+1}\rangle}{\langle\text{Swap}_{A}^{j}\rangle} & =\frac{\sum_{\bm{\sigma}_{1}}\sum_{\bm{\sigma}_{2}}\psi_{\bm{\lambda}}(\bm{\sigma}_{1})\psi_{\bm{\lambda}}(\bm{\sigma}_{2})\psi_{\bm{\lambda}}(\bm{\sigma}_{12}^{j+1})\psi_{\bm{\lambda}}(\bm{\sigma}_{21}^{j+1})}{\sum_{\bm{\sigma}_{1}}\sum_{\bm{\sigma}_{2}}\psi_{\bm{\lambda}}(\bm{\sigma}_{1})\psi_{\bm{\lambda}}(\bm{\sigma}_{2})\psi_{\bm{\lambda}}(\bm{\sigma}_{12}^{j})\psi_{\bm{\lambda}}(\bm{\sigma}_{21}^{j})}\\
 & =\frac{\sum_{\bm{\sigma}_{1}}\sum_{\bm{\sigma}_{2}}\psi_{\bm{\lambda}}(\bm{\sigma}_{1})\psi_{\bm{\lambda}}(\bm{\sigma}_{2})\psi_{\bm{\lambda}}(\bm{\sigma}_{12}^{j})\psi_{\bm{\lambda}}(\bm{\sigma}_{21}^{j})\frac{\psi_{\bm{\lambda}}(\bm{\sigma}_{12}^{j+1})\psi_{\bm{\lambda}}(\bm{\sigma}_{21}^{j+1})}{\psi_{\bm{\lambda}}(\bm{\sigma}_{12}^{j})\psi_{\bm{\lambda}}(\bm{\sigma}_{21}^{j})}}{\sum_{\bm{\sigma}_{1}}\sum_{\bm{\sigma}_{2}}\psi_{\bm{\lambda}}(\bm{\sigma}_{1})\psi_{\bm{\lambda}}(\bm{\sigma}_{2})\psi_{\bm{\lambda}}(\bm{\sigma}_{12}^{j})\psi_{\bm{\lambda}}(\bm{\sigma}_{21}^{j})}\\
 & =\frac{\sum_{\bm{\sigma}_{1}}\sum_{\bm{\sigma}_{2}}P^{j}(\bm{\sigma}_{1},\bm{\sigma}_{2})\mathcal{R}^{j}(\bm{\sigma}_{1},\bm{\sigma}_{2})}{\sum_{\bm{\sigma}_{1}}\sum_{\bm{\sigma}_{2}}P^{j}(\bm{\sigma}_{1},\bm{\sigma}_{2})}\\
 & =\langle\mathcal{R}^{j}(\bm{\sigma}_{1},\bm{\sigma}_{2})\rangle_{P},
\end{split}
\end{equation}
where we defined the probability distribution $P^{j}(\bm{\sigma}_{1},\bm{\sigma}_{2})=\psi_{\bm{\lambda}}(\bm{\sigma}_{1})\psi_{\bm{\lambda}}(\bm{\sigma}_{2})\psi_{\bm{\lambda}}(\bm{\sigma}_{12}^{j})\psi_{\bm{\lambda}}(\bm{\sigma}_{21}^{j})$
and the observable 
\begin{equation}
\mathcal{R}^{j}(\bm{\sigma}_{1},\bm{\sigma}_{2})=\frac{\psi_{\bm{\lambda}}(\bm{\sigma}_{12}^{j+1})\psi_{\bm{\lambda}}(\bm{\sigma}_{21}^{j+1})}{\psi_{\bm{\lambda}}(\bm{\sigma}_{12}^{j})\psi_{\bm{\lambda}}(\bm{\sigma}_{21}^{j})}.
\end{equation}
To compute the expectation value $\langle\mathcal{R}^{j}(\bm{\sigma}_{1},\bm{\sigma}_{2})\rangle_{P^{j}}$
we employ standard Monte Carlo simulation, where spin configurations
$(\bm{\sigma}_{1},\bm{\sigma}_{2})$ for the two copies are sampled
from the probability distribution $P^{j}(\bm{\sigma}_{1},\bm{\sigma}_{2})$.
To compute the entanglement entropy for a half-chain we run $N/2$
separate Markov chain for each $j=1,\dots,N/2$ and compute the entropy
as 
\begin{equation}
S_{2}(\rho_{N/2})=-\sum_{j=0}^{N/2-1}\log\langle\mathcal{R}^{j}(\bm{\sigma_{1}},\bm{\sigma_{2}})\rangle_{\textrm{mc}}.
\end{equation}

\section{Overfitting}
\begin{figure}[b]
\noindent \centering{}\includegraphics[width=0.75\columnwidth]{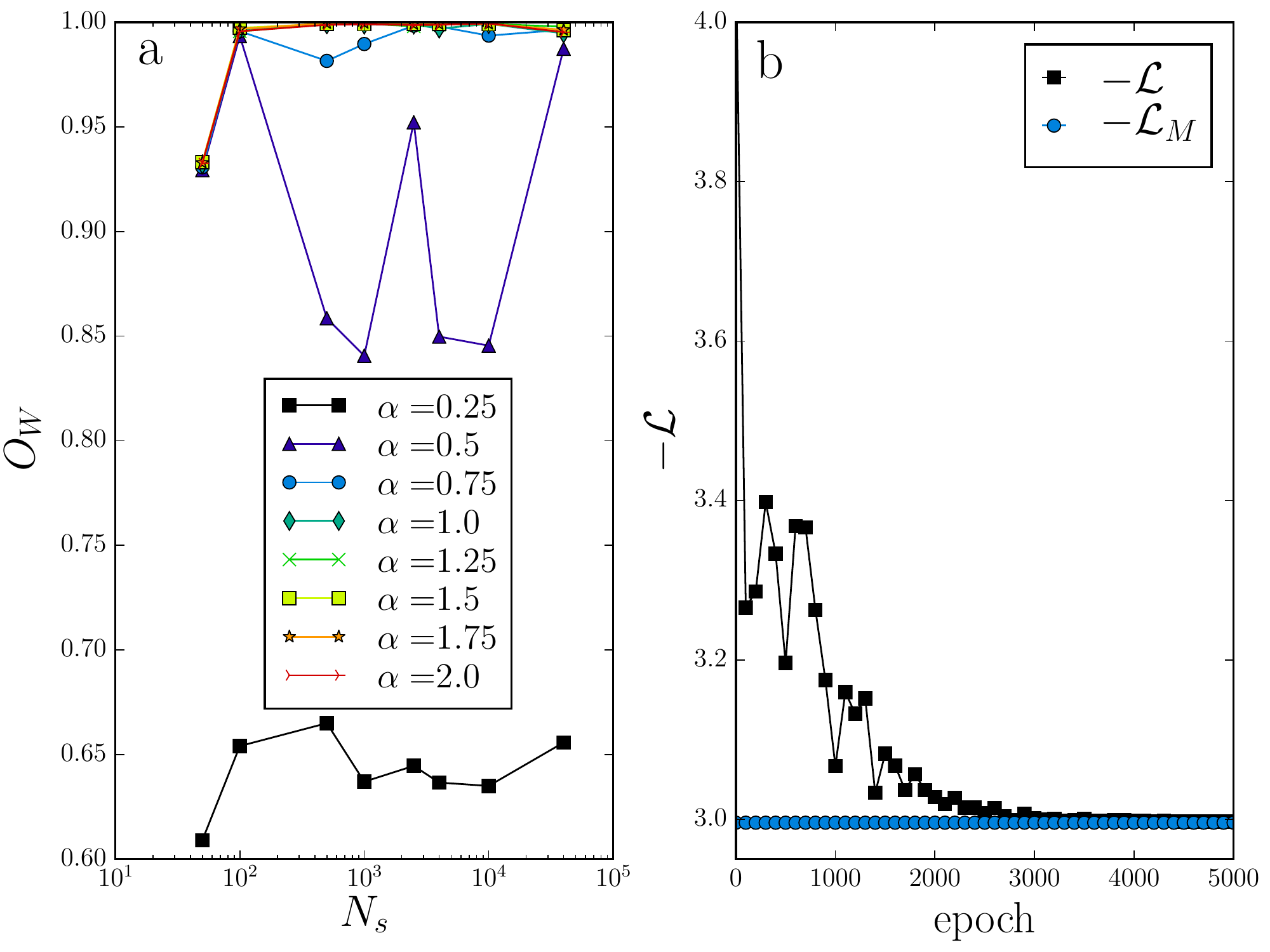}
\caption{Investigating overfitting of RMBs in the learning of the W state.
(a) The overlap $O$ between the W state wave-function and the RBM
wave-function as a function of the size of the training data $N_{s}$
for different values of $\alpha$ and fixed value of $N=20$. (b)
The negative log-likelihood measure of a held-out test set with $N_{t}=10000$
as a function of the training epoch for an RBM with $N=20$ and $\alpha=1$
trained on a dataset with $N_{s}=40000$.}
\label{overf} 
\end{figure}
As per all machine learning applications, the training process should
be carefully designed to avoid overfitting. This issue occurs when
a complex model does not generalize well to unseen data, even though
the model fits well the training data. In our experiments with RBMs,
overfitting may manifest itself when the model is excessively powerful,
i.e. when $\alpha\gg1$, and/or when the data sets used during the
training stage are statistically small. We investigate the overfitting
in the training of our RBMs applied to the $W$ state in two ways.
First, we track the overlap $O$ between the W state wave-function
and the RBM wave-function, which should approach 1 for a properly
trained model. In Fig.~\ref{overf}(a) we present the overlap $O$
between the W state wave-function and the RBM wave-function as a function
of the size of the training data $N_{s}$ for different values of
$\alpha$ and fixed number of qubits $N=20$. We first note that for
small $\alpha$ the RBM states are generically poor approximations
to the W state. As long as the size of the training dataset is small,
increasing $\alpha$ is not enough to achieve a significant improvement
in the overlap $O$. Upon increasing $N_{s}$, however, increasing
the capacity of the RBM to $\alpha=1$ results in overlaps approaching
$O\approx1$. Crucially, further increasing $\alpha$ does not deteriorate
the values of $O$ attained by the RBMs, and it rather saturates,
which we attribute to having significantly large datasets that prevent
overfitting. Second, we track the probability of a held-out test set
during the training, which relates to the objective function that
we minimize during the training. Given a dataset, the log-likelihood
of the data is given by 
\begin{equation}
\mathcal{L}=\frac{1}{|D_{b}|}\sum_{\bm{\sigma_{j}}\in D_{b}}\log\left|\frac{p_{\bm{\lambda}}(\bm{\sigma}_{j})}{{Z_{\bm{\lambda}}}}\right|.
\end{equation}

Notice that because the calculation of $\mathcal{L}$ requires the
evaluation of intractable partition functions $Z_{\bm{\lambda}}$,
we restrict our calculation to small systems with $N=20$ and $\alpha=1$,
where an exact evaluation of $Z_{\bm{\lambda}}$ is still possible.
For a generic evaluation of $Z_{\bm{\lambda}}$, one has to resort
to advanced sampling techniques such as parallel tempering and annealed
importance sampling.\cite{SalMurray08} From this perspective, overfitting
would be evidenced by a continuous improvement of both the training
and held-out $\mathcal{L}$ followed by degradation of \textit{only}
the held-out $\mathcal{L}$ due to the excessive adjustment of the
parameters of the RBM that improves the $\mathcal{L}$ of the training
dataset exclusively. In Fig.~\ref{overf}(b) we display the evolution
of the held-out $\mathcal{L}$ as the training progresses. We also
report the value of the held-out data set $\mathcal{L}_{M}=-\log N$
under the distribution it came from, i.e., the modulus square of the
amplitudes in the W state, which is the optimal value the RBM would
achieve if it perfectly described the data. Apart from noticing that
no evidence of overfitting is found, we emphasize that the held-out
$\mathcal{L}$ approaches the theoretical value $\mathcal{L}_{M}$
near the end of the training, which means that the RBM describes the
distribution of the data remarkably well.

\end{document}